\documentclass[nature,preprint,onecolumn,superscriptaddress,floatfix]{revtex4-1}

\usepackage{epsfig,dcolumn,amsmath,latexsym}
\usepackage{graphicx}
\usepackage{subfigure}
\usepackage[normalem]{ulem}
\usepackage{cancel}
\usepackage{multirow}
\usepackage{longtable}

\begin{document}

\title{Tunable electronic structure and topological properties of LnPn (Ln=Ce, Pr, Sm, Gd, Yb; Pn=Sb, Bi)}




\author{Xu Duan}
 \affiliation{Condensed Matter Group,
  Department of Physics, Hangzhou Normal University, Hangzhou 311121, China}
\author{Fan Wu}
 \affiliation{Center of Correlated Materials, Zhejiang University, Hangzhou 310058, China}
 \affiliation{Department of Physics, Zhejiang University, Hangzhou 310027, China}
\author{Jia Chen}
 \affiliation{Condensed Matter Group,
  Department of Physics, Hangzhou Normal University, Hangzhou 311121, China}
 \affiliation{Department of Physics, Zhejiang University, Hangzhou 310027, China}

\author{Peiran Zhang}
 \affiliation{Center of Correlated Materials, Zhejiang University, Hangzhou 310058, China}
 \affiliation{Department of Physics, Zhejiang University, Hangzhou 310027, China}
\author{Yang Liu}
\author{Huiqiu Yuan}
 \affiliation{Center of Correlated Materials, Zhejiang University, Hangzhou 310058, China}
 \affiliation{Department of Physics, Zhejiang University, Hangzhou 310027, China}
\author{Chao Cao}
 \affiliation{Condensed Matter Group,
  Department of Physics, Hangzhou Normal University, Hangzhou 311121, China}


\begin{abstract}
Recently, the rare-earth monopnictide compounds LnPn have attracted considerable attention in condensed matter physics studies due to their possible topological properties. We have performed systematic first principles study of the electronic structure and band topology properties of LnPn (Ln=Ce, Pr, Sm, Gd, Yb; Pn=Sb, Bi). Assuming the $f$-electrons are well localized in these materials, both hybrid functional and modified Becke-Johnson calculations yield electronic structure in good agreement with experimental observations, while generalized gradient approximation calculations severely overestimate the band inversions. From Ce to Yb, a systematic reduction of band inversion with respect to the increasing Ln atomic number is observed, and $\mathcal{Z}_2$ for CePn and YbPn are [1;000] and [0;000], respectively. In both hybrid functional and modified Becke-Johns calculations, a topologically nontrivial to trivial transition is expected around SmSb for the antimonides and around DyBi for the bismuthides. Such variation is related with lanthanide contraction, but is different from simple pressure effects.
\end{abstract}

\maketitle

\section*{Introduction}

The rare-earth monopnictide compounds were discovered decades ago, including other members in the same family (e.g. CeN, CeP, etc). All the compounds crystallize in simple rocksalt face-centered cubic structure except YbBi, but they exhibit a diversity of physical properties. Most of these compounds are antiferromagnetic at low temperature\cite{PhysRevB.49.15096,PhysRevB.93.205152,LnBi.Magnetism.1965,LnBi.Mag}, except for the nonmagnetic yittrium and lanthanum compounds, as well as the praseodymium compounds that are Van Vleck type paramagnetic\cite{PrPn.Magnetism}. Angular resolved photoemission spectroscopy (ARPES) and de Haas van Alphen (dHvA) measurements have been performed\cite{KASUYA19879,PhysRevB.56.13654,PhysRevB.58.7675,PhysRevB.54.9341,PhysRevB.33.6730,MORITA1997192,OZEKI1991499,CeSb.Heavy.Hole.dHvA,PhysRevB.96.125122} to probe the electronic structure of these compounds. Both dynamical mean-field theory (DMFT) and LDA+$U$ calculations have suggested that the 4$f$-electrons are almost fully localized in Pn=Sb/Bi compounds\cite{PhysRevLett.79.2546,PhysRevB.86.115116,PhysRevB.74.085108,CePn.Pressure.LDAU}, especially for the bismuth compounds. For CePn, the Ce-4$f$ orbitals slightly mixes with the Ce-5$d$ orbitals ($d$-$f$ mixing) and Pn-$p$ orbitals ($p$-$f$ mixing)\cite{PhysRevLett.88.207201,KASUYA19879}, and the former leads to Kondo-like behaviors in this low density carrier system, which almost vanishes for CeBi\cite{PhysRevB.86.115116}.

In addition, LaPn system was proposed to be a topological system by first principles calculation\cite{2015arXiv150403492Z}. Although the initial proposal argued that the LaN is a 3D Dirac-semimetal and all other LaPn are topological insulators, it was later argued that the band inversion was severely overestimated in the calculation, thus only LaBi is topologically nontrivial in the series\cite{PhysRevB.93.235142}. Both ARPES and transport measurements have confirmed that LaSb is a topologically trivial electron-hole compensated semimetal\cite{PhysRevLett.117.127204} similar to YSb\cite{PhysRevLett.117.267201}; while LaBi is topologically nontrivial with clear evidence of protected surface states\cite{PhysRevB.93.241106,PhysRevB.95.115140,Nayak:2017aa,PhysRevB.94.165163,PhysRevB.96.161103}. It has also been proposed from first-principles method that topological transition from trivial to nontrivial can be realized in LaSb by applying hydraulic pressure\cite{PhysRevB.96.081112}. More recently, ARPES measurements have been performed with CeSb and CeBi compounds to investigate their topological properties\cite{PhysRevB.96.035134,PhysRevB.96.041120,2016arXiv160408571A,PhysRevLett.120.086402,PhysRevB.98.085103,REBi.Liu.ARPES}, and transport evidence for possible Weyl fermion has been found in CeSb at ferromagnetic state\cite{Guo:2017aa}. Large magnetoresistance has also been reported in NdSb\cite{PhysRevB.93.205152}. Therefore, the renewed interests in LnPn system focus on possible topological properties of these materials. However, a systematic first-principles study of rare-earth monopnictide compound in this aspect is still missing.

In this article, we present our latest first-principles study of LnPn (Ln=Ce, Pr, Sm, Gd, Yb; Pn=Sb, Bi). In particular, we compare the electronic structure obtained using Perdew-Burke-Ernzerhof (PBE), modified-Becke-Johnson (mBJ), and Heyd-Scuseria-Ernzerhof hybrid (HSE06) functionals, as well as the band inversion features and the associated $\mathcal{Z}_2$ indices. We find good agreement between the mBJ/HSE06 results and quantum oscillation measurements, while PBE results usually yields substantially larger $\beta$ and $\gamma$ band frequencies due to overestimation of band inversions. From Ce to Yb, a systematic reduction of band inversion related with lanthanide contraction is observed, and the topologically nontrivial to trivial transition takes place around SmSb for antimonides and DyBi for bismuthides, respectively. Finally, in contrast to lanthanide contraction, simple pressure effects will increase the band inversion gap.


\section*{Results}

\begin{figure}[h]
\includegraphics[width=15cm]{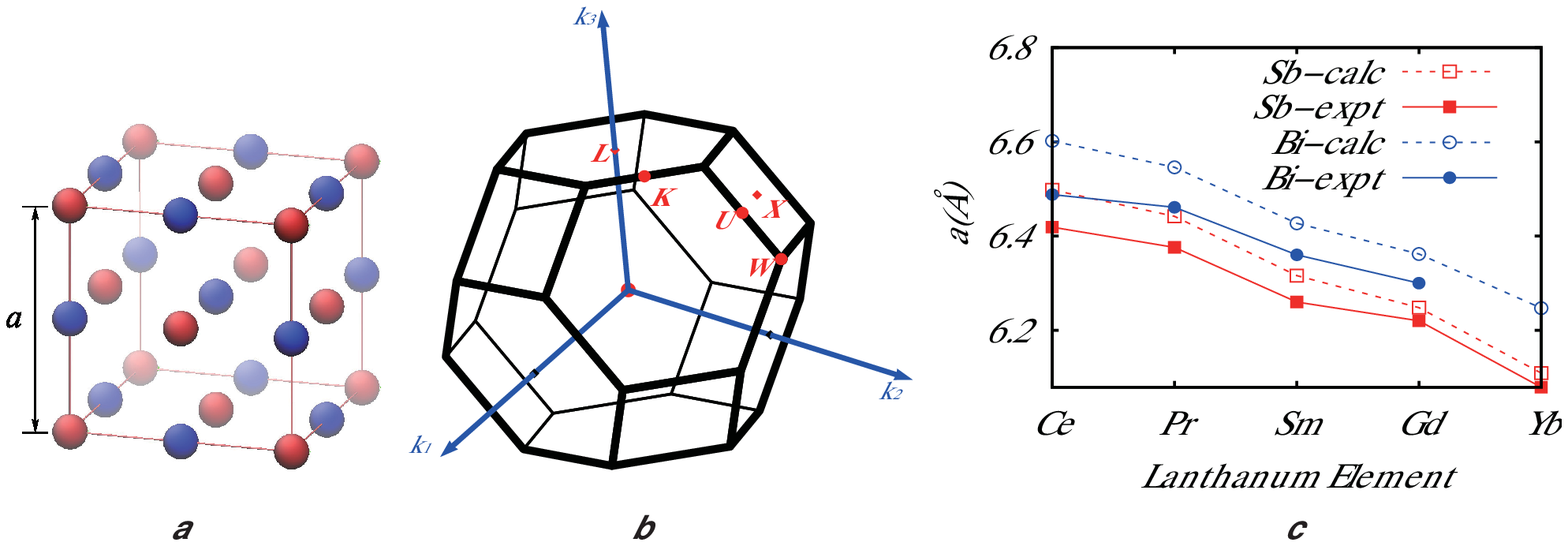}
 \caption{Crystal structure and first Brillouine zone of LnPn. {\bf a} Crystal structure of LnPn. The red atoms are lanthanide elements while blue atoms are pnictogen. {\bf b} The first Brillouine zone of LnPn and its high symmetry points. {\bf c} Comparison between the lattice constants obtained from density functional theory (DFT) calculations (solid lines, empty square/circle) and experiments (dashed lines, solid square/circle).\label{fig:geo}}
\end{figure}

We show the crystal structure and its first Brillouine zone of LnPn in fig \ref{fig:geo}a and \ref{fig:geo}b, respectively. In fig. \ref{fig:geo}c, we compare the lattice constants from structural relaxation with those obtained in experiments. The calculated lattice constants are slightly larger than the experimental observation, since generalized gradient approximation (GGA) tends to underestimate binding. Nevertheless, all the calculated values are within 3\% errorbar of density functional theory (DFT) calculations. In both calculation and experiments, the lattice constants of either bismuth compounds or antimony compounds decrease as the lanthanide element moves towards the end of the periodic table, i.e. the lanthanide contraction. Overall, the calculations correctly reproduce the trend of lattice constant change as the lanthanide element changes, demonstrating the validity of the DFT calculation. It is worth noting that to the best of our knowledge, there is no report on the rock-salt structured YbBi compound in the literature, possibly due to the small Yb$^{3+}$ radius.

\subsection*{Electronic band structure}

\begin{figure}[h]
\includegraphics[width=15cm]{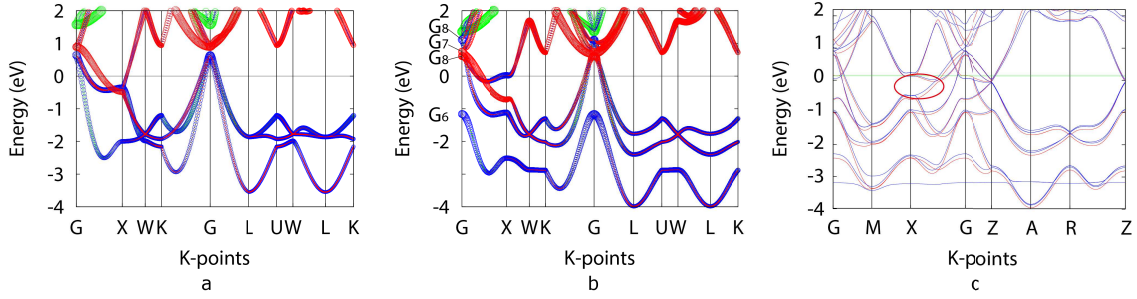}
  \caption{General features of LnPn band structure. {\bf a-b} Electronic band structure of CeBi in nonmagnetic state {\bf a} without and {\bf b} with spin-orbit coupling. The orbital character weights are represented by the size of the circles. The blue circles are Bi-6$p$ orbitals, red circles are Ce-5$d$ t$_{2g}$ orbitals, and green circles are Ce-5$d$ e$_g$ orbitals. {\bf c} Comparison between $f$-core nonmagnetic state CeBi band structure in the folded Brillouin zone (red lines) and LDA+$U$ antiferromagnetic (AFM) state CeBi band structure (blue lines). Notice that in AFM state, the Brillouin zone is folded, so the band structure is more complicated. Nevertheless, the inversion gap is still present and the anti-crossing feature can be identified along $\Gamma$-$X$ (red-circled area). \label{fig:bs_cebi_pbe}}
 \end{figure}

To investigate the electronic structure of LnPn compounds, we first take the nonmagnetic band structure of CeBi to study some general features. Without spin-orbit coupling, the electron states near the Fermi level are dominated by the pnictogen $p$-orbitals and lanthanide t$_{2g}$-orbitals (fig. \ref{fig:bs_cebi_pbe}a). Normally, the energy level of lanthanide t$_{2g}$-orbitals are higher than that of the pnictogen $p$-orbitals. However, such order can be reversed after considering the spin-orbit coupling in some of these LnPn compounds at X ($\pi$, $\pi$, 0), leading to a band crossing below $E_F$ along $\Gamma$-X (fig. \ref{fig:bs_cebi_pbe}b). With the spin-orbit coupling, the 3 degenerate $p$-orbitals further split into a doubly degenerate $\Gamma_8$ state and a $\Gamma_6$ state at $\Gamma$ point; while the 3 degenerate t$_{2g}$-orbitals split into a doubly degenerate $\Gamma_8$ state and a $\Gamma_7$ state at $\Gamma$. In addition, the spin-orbit coupling will gap out the band crossing between $\Gamma$-X due to the band inversion. As a result, one can define a partial gap at the electron occupation in the whole Brillouin zone for these materials. Therefore, the $\mathcal{Z}_2$ indices are well defined for these materials. With spin-orbit coupling, there are 3 bands crossing the Fermi level, among which two of them are hole-like near $\Gamma$ and another electron-like one near X point. Since the electronic states are close to $E_F$ only around $\Gamma$ and X, and the band topology is related with the anti-crossing between $\Gamma$-X, we shall focus on the band dispersion along this direction.

We have also performed calculations with itinerant $f$-electron in both nonmagnetic state and antiferromagnetic state for CeBi. By taking the $f$-orbitals into valence, they become very itinerant and dominate the density of states near $E_F$. This is a known artifact of local-density approximation due to lack of proper electron-correlation effect when it is applied to open-shell $f$ compounds. To obtain the correct band structure, we have applied LDA+$U$ method with $U_f$=6.1 eV in the antiferromagnetic calculations (fig. \ref{fig:bs_cebi_pbe}c). The resulting band structure is very close to the nonmagnetic band structure we obtained with $f$-core calculations, except that the antiferromagnetic band structure is folded (since the antiferromagnetic Brillouin zone is half of the original one), and that an extra flat $f$-band appear at approximately 3 eV below $E_F$. Both the inversion gap and the  anti-crossing feature are present, and the size of the inversion gap is similar. 

\begin{figure}[h]
\includegraphics[width=15cm]{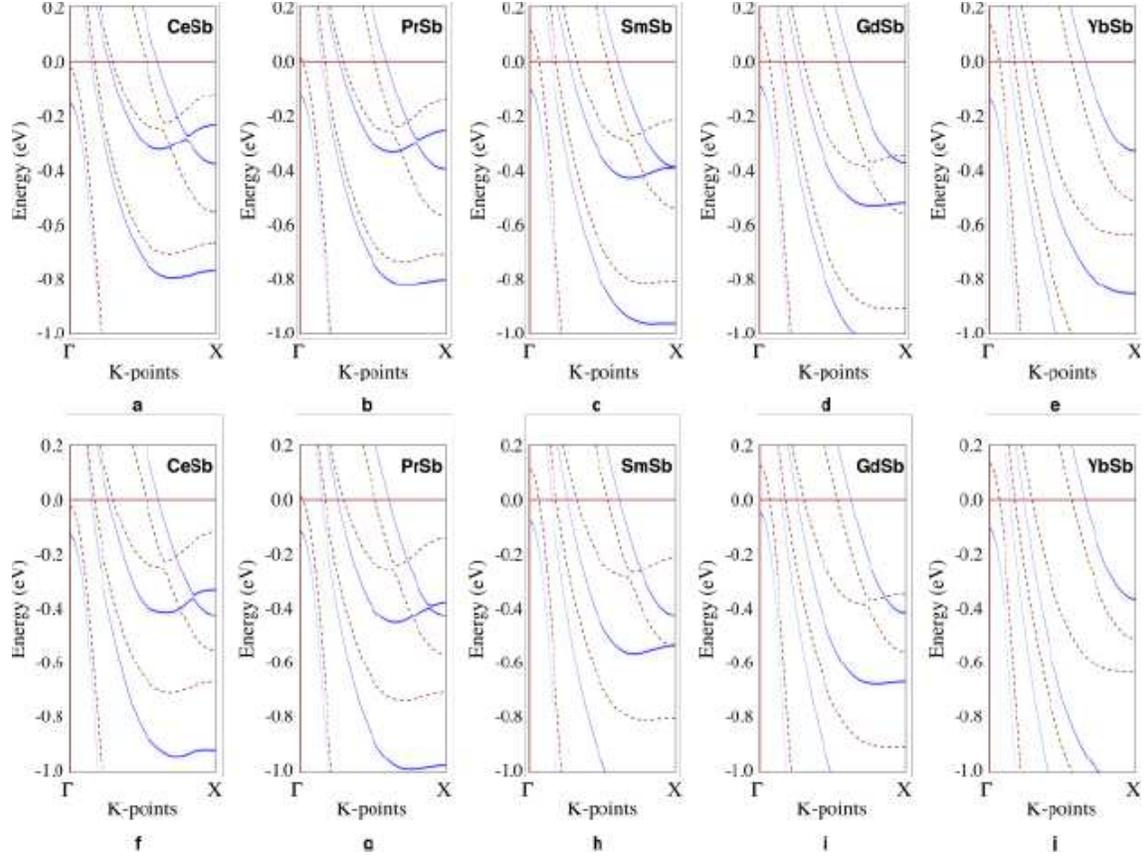}
 \caption{Electronic band structure of LnSb along $\Gamma$-$X$. The dashed lines are band structure obtained with PBE exchange-correlation functionals, whereas the blue solid lines are obtained using {\bf a-e} mBJ potential and {\bf f-j} HSE06 hybrid functionals, respectively. From left to right are {\bf a}\&{\bf f} CeBi, {\bf b}\&{\bf g} PrBi, {\bf c}\&{\bf h} SmBi, {\bf d}\&{\bf i} GdBi, and {\bf e}\&{\bf j} YbBi, respectively.\label{fig:bs_xsb}}
\end{figure}

In fig.\ref{fig:bs_xsb}, we show the electronic band structures of LnSb compounds. For CeSb, the band inversion occurs for t$_{2g}$ orbitals and $p$ orbitals, resulting in a large inversion gap of $\approx$400 meV at X under PBE approximation. However, since PBE band energies are not quasi-particle energies, and that PBE band widths are overestimated, the inverted gap sizes in PBE are also overestimated. Therefore, we have also performed calculations in mBJ and HSE06 hybrid functional as well. Due to the overwhelming amount of calculation load for HSE06, the band structures of HSE06 hybrid functionals were obtained by Wannier fitting. The inversion gap size is substantially reduced (142 meV in mBJ and 92 meV in HSE06), but the band anticrossing feature remains. From CeSb to YbSb, the inversion gap size reduces with respect to the increasing atomic number of the lanthanide element, and eventually disappears for YbSb for all functionals. For the mBJ method, the transition occurs at SmSb, where the Sm t$_{2g}$ orbitals and Sb $p$-orbitals are nearly degenerate ( $<1$ meV). For the HSE06 functional, the inversion gap sizes are generally smaller than those in mBJ calculations, and therefore this transition occurs closer to PrSb. While the decreasing trend of band inversion gap with heavier rare-earth element is in line with the ARPES results, quantitative deviations are still present in LnSb compounds. For example, ARPES measurements have shown that both CeSb and PrSb are close to the boundary of topologically trivial-nontrivial transitions, while SmSb is clearly on the trivial side with sizable non-inverted gap. \cite{REBi.Liu.ARPES,PhysRevLett.120.086402}.

\begin{figure}[h]
\includegraphics[width=15cm]{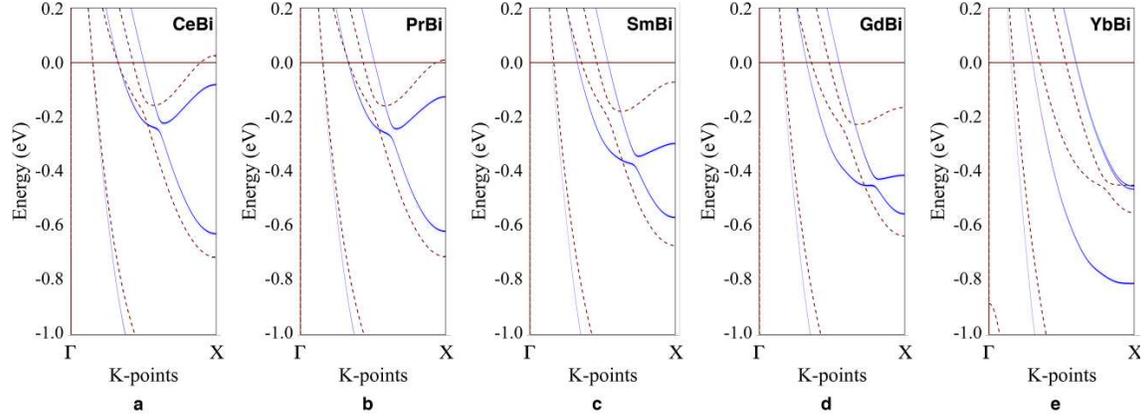}
  \caption{Electronic band structure of LnBi along $\Gamma$-$X$. The dashed lines are band structure obtained with PBE exchange-correlation functionals, whereas the blue solid lines are obtained using HSE06 hybrid functionals, respectively. The mBJ results are very similar to the HSE06 results. From left to right are {\bf a} CeBi, {\bf b} PrBi, {\bf c} SmBi, {\bf d} GdBi, and {\bf e} YbBi, respectively. \label{fig:bs_xbi}}
\end{figure}

Similar variation of inversion gap with respect to the increasing atomic number is also present in the bismuth compounds (fig. \ref{fig:bs_xbi}). Due to the increased spin-orbit coupling in bismuth, the inversion gap is substantially enlarged. Under PBE approximation, the band inversion was so strong that the band from $p$-orbitals would cross the Fermi level around X in CeBi and PrBi, creating an additional hole pocket around X. However, this hole pocket is absent in either mBJ or HSE06 calculations. Using the HSE06 functional, the gap size is 550 meV in CeBi, 495 meV in PrBi, 270 meV in SmBi, 140 meV in GdBi and eventually disappears in YbBi. Therefore, a transition from anticrossing to non-crossing band also exists in the bismuth compounds, presumably between GdBi and YbBi. In fact, we have located the position of this transition closest to DyBi, where the direct gap between Dy-t$_{2g}$ and Bi-6$p$ orbitals is $\approx$ 8 meV (5 meV) in mBJ (HSE06) calculations. 

In addition to the variation of inversion gap, the energy level of the $\Gamma_6$ state at $\Gamma$ point exhibits some systematic changes. Under PBE approximation, the $\Gamma_6$  state is slightly below $E_F$ in CeSb and slightly above $E_F$ in PrSb. It rises as one moves towards YbSb. Therefore, an additional hole pocket would appear and gradually increases as one increases the atomic number under PBE. However, for both mBJ and HSE06 calculations, the $\Gamma_6$ state is much lower than it is in PBE, and is always below $E_F$. Nevertheless, the trend remains visible, and the $\Gamma_6$ state in GdSb is much closer to $E_F$ than it is in CeSb. Such change in $\Gamma_6$ state is also present in bismuth compounds. However, the $\Gamma_6$ state is more than 1 eV below $E_F$ in these compounds, and therefore are irrelevant. Both mBJ and HSE06 results agree very well with ARPES measurements for CeBi, PrBi, SmBi and GdBi\cite{PhysRevB.98.085103}.

Using the maximally projected Wannier function method, we have also calculated the angular dependence of dHvA frequencies of these materials\cite{ROURKE2012324} (TAB. \ref{tab:dhva}) and compared with available experimental observed values. It is apparent that PBE considerably overestimates all frequencies; while the mBJ and HSE06 results agree well with the experimental observations at least for $\alpha$ and $\gamma$ bands. Noticing that these bands are indeed the electron-type pockets due to t$_{2g}$ orbitals close to $X$, and are directly relevant to the band inversion features. Moreover, the frequencies for $\beta$ bands obtained in mBJ and HSE06 calculations also agree with experimental values except for CeSb and PrSb. It is worthy noting that the $\beta$ bands are hole-type pockets due to $p$-orbitals close to $\Gamma$, and are very heavy in CeSb (renormalized $\approx$ 10 times for $\beta_4$). In short, the assumption of localized $f$-electron appear to be quite rational for these compounds.


\subsection*{Topological indices and surface State}

The crystal structure of LnPn compounds are centrosymmetric, therefore the $\mathcal{Z}_2$ indices can be determined by calculating the band parities at 8 time reversal invariant momenta (TRIM), namely the $\Gamma$ point, 3 $X$ points, and 4 $L$ points. For CePn, PrPn, SmBi and GdBi, the resulting $\mathcal{Z}_2$ are nontrivial [1;000], and are consistent in PBE, mBJ and HSE06 calculations. For YbSb, all calculations yields trivial [0;000]. For SmSb, GdSb, DyBi and YbBi, $\mathcal{Z}_2$ is nontrivial [1;000] in PBE calculations, but [0;000] in both mBJ and HSE06 calculations. These results agree with our band structure analysis in previous section. In particular, we note that SmSb and DyBi are on the edge to the transition from topologically nontrivial to trivial in more accurate mBJ and HSE06 calculations. 

\begin{figure}[h]
\includegraphics[width=15cm]{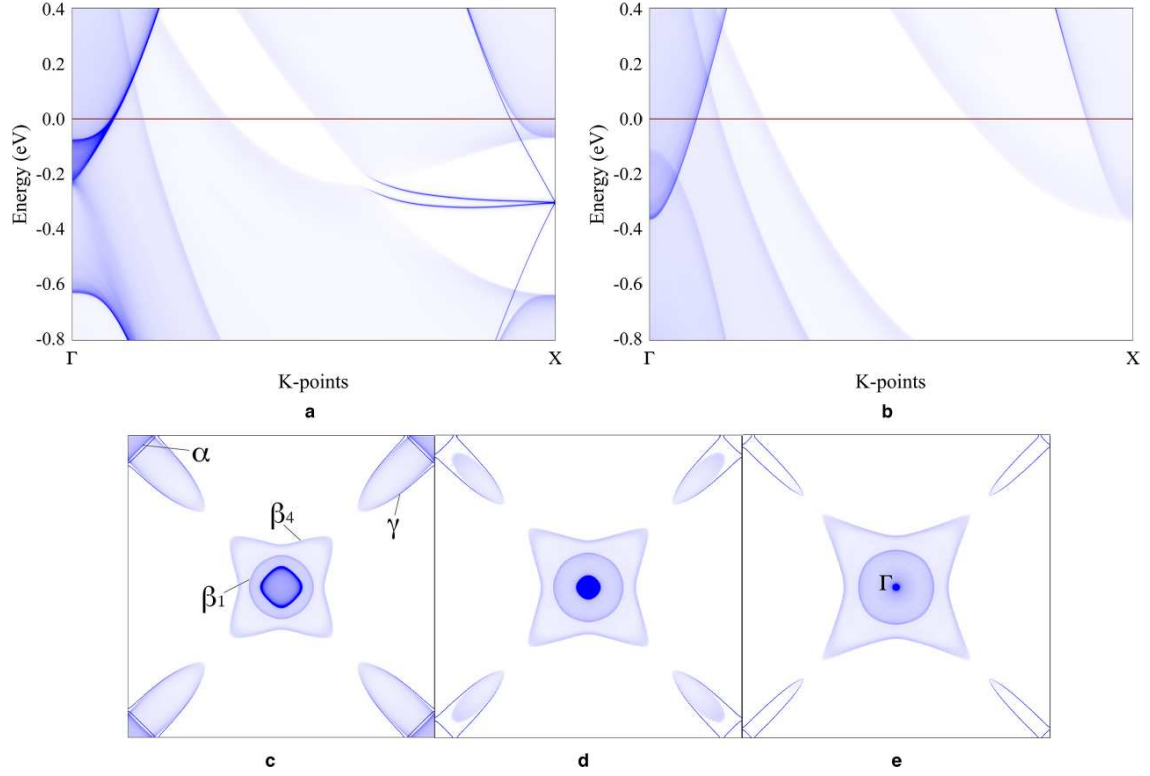}
  \caption{Surface state of LnPn. {\bf a-b} Band structure of a semi-infinite [001] surface of {\bf a} topologically nontrivial CeBi and {\bf b} topologically trivial YbSb using HSE06. {\bf c-e} Surface state of semi-infinite [001] surface of CeBi using HSE06 at {\bf c} $E_F$, {\bf d} $E_F-0.1$ eV, and {\bf e} $E_F-0.2$ eV. The zone center is $\Gamma$, whereas the corners are $X$. The bulk states can be identified with continuum-like spectrum, while the surface states appear as individual lines. In {\bf c}, the $\alpha$, $\beta$, $\gamma_1$, and $\gamma_4$ pockets are indicated. \label{fig:surf_xpn}}
\end{figure}

A crucial difference between topological trivial and non-trivial state lies at their boundary, where topologically protected surface state exists for the non-trivial state. Employing the surface Green's function method, we have obtained the band structure and surface state for the semi-infinite [001] surface of LnPn compounds. In fig. \ref{fig:surf_xpn} (a) and (b), we compare two typical cases CeBi and YbSb for nontrivial and trivial case, respectively. The topological surface states are prominent in CeBi (fig. \ref{fig:surf_xpn}a), where 2 Dirac-cones due to these states are formed at X point. The two Dirac points are very close in energy, mainly because the valence difference between surface atoms and bulk atoms are ignored in our surface Green's function calculations. Nevertheless, these two points are not energetically degenerate, and they will further separate if the valence difference is taken into consideration (e.g. in slab model calculations\cite{PhysRevB.98.085103}). However, slab model calculation with mBJ is flawed due to technical reasons, and HSE06 is extremely expensive and beyond our current calculation capability. In contrast, only bulk states are visible in YbSb compound (fig. \ref{fig:surf_xpn}b). We have also performed calculations of the CeBi [001] surface at different energy [fig. \ref{fig:surf_xpn}(c-e)]. At the zone center ($\Gamma$), only bulk state is visible at all energies. The surface states can be observed around the zone corner $\bar{\mathrm{X}}$. At $E_F$, the surface states are mixed with bulk states, and cannot be easily distinguished. Below $E_F-0.1$ eV, the bulk states are void in the band inversion gap, and the surface states become prominent. At $E_F-0.2$ eV, the surface states dominates the zone corner. 

\begin{figure}[h]
\includegraphics[width=15cm]{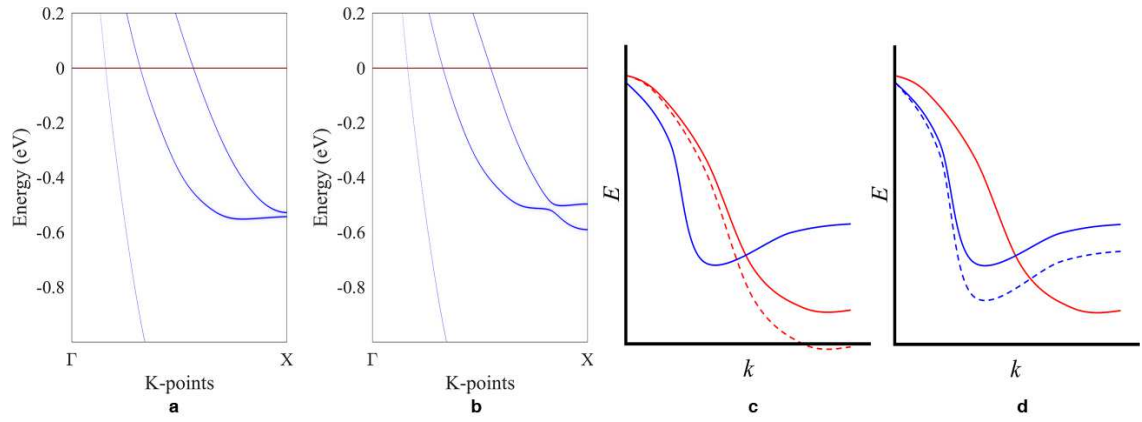}
  \caption{Effect of pressure. HSE06 band structure between $\Gamma$ and $X$ of DyBi at {\bf a} ambient pressure ($a$=6.31\AA) and {\bf b} 2.1 GPa ($a$=6.25\AA). {\bf c-d} are schematic demonstration of enhanced dispersion of {\bf c} Ln-t$_{2g}$ band and {\bf d} Pn-$p$ band, respectively. \label{fig:bs_dybi}}
\end{figure}

Finally, we discuss the inverted gap and $\mathcal{Z}_2$ variations with respect to the rare-earth atomic numbers. In LnPn, there are two effects due to replacing the rare-earth element. Firstly, when $X$ is replaced with heavier rare-earth element, the spin-orbit coupling is increased. Since $\Delta\propto Z_X^2$ where $Z_X$ is the atomic number of $X$, the percentage increase of $\Delta$ is proportional to $2\vert Z_X-Z_{\mathrm{Ce}}\vert / Z_{\mathrm{Ce}}$\cite{PhysRevB.90.165108}. In fact, the t$_{2g}$ splitting is 0.12 eV in CeSb, 0.15 eV in SmSb and 0.18 eV in YbSb,  roughly following the estimation. Secondly, when Ln is replaced, it causes lanthanide contraction. However, it is not equivalent to chemical pressure effect. In order to demonstrate this, we show the band structure of DyBi along $\Gamma$-$X$ at ambient pressure (6.31\AA), as well as with YbBi lattice constant (6.25\AA, or equivalent to 2.1 GPa) (fig. \ref{fig:bs_dybi}). It is apparent that DyBi at ambient pressure is topologically trivial (also confirmed by $\mathcal{Z}_2$ calculation), whereas it is nontrivial at 2.1 GPa ($\mathcal{Z}_2$ is [1;000]). It is worthy noting that similar pressure-induced topological transition has also been proposed for LaSb between 3 $\sim$ 4 GPa\cite{PhysRevB.96.081112}. The reduction of lattice constant increases the hopping between orbitals, and therefore band dispersion. In LnPn, the increase of Ln-t$_{2g}$ band dispersion enhances the inversion gap (fig. \ref{fig:bs_dybi}c) while the increase of Pn-$p$ band dispersion reduces the gap (fig. \ref{fig:bs_dybi}d). In fact, the largest nearest neighboring hopping between Bi-6$p$ orbitals increases 77\% (from 0.13 eV in CeBi to 0.23 eV in YbBi); while the largest nearest neighboring hopping between Ln-t$_{2g}$ orbitals increases only  5\% (from 0.43 eV in CeBi to 0.45 eV in YbBi). This in turn reflects the dispersion changes, which suggests that $p$-band dispersion is more sensitive to the lanthanide contraction than $d$-band. On the contrary, when a pure pressure is applied, the largest nearest neighboring hopping between Ln-t$_{2g}$ orbitals increases much faster than that between Pn-$p$ orbitals, while the spin-orbit coupling strength $\Delta$ remains relatively constant. For example, if we compress that lattice constant of DyBi by 5\%, the largest nearest neighboring hopping between Dy-t$_{2g}$ orbitals increases 20\% (from 0.44 eV to 0.53 eV); whereas the largest nearest neighboring hopping between Bi-6$p$ orbitals only increases 10\% (from 0.19 eV to 0.22 eV). In short, the inverted gap in LnPn systems increases with respect to increasing spin-orbit coupling strength $\Delta$, increasing Ln-t$_{2g}$ band dispersion, or decreasing Pn-$p$ band dispersion. 

\section*{Discussion}
In conclusion, we have studied the electronic structure and band topology of LnPn compounds using state-of-art density-functionals. While PBE overestimates the band inversion gap, both mBJ and HSE06 functionals yield similar results. From CePn to YbPn, the atomic number increases but the band inversion gap size reduces. Therefore, a topological nontrivial to trivial transition is expected between PrSb and SmSb in LnSb compounds; while for LnBi compounds, the transition is expected around DyBi. Such variation in the inversion gap is related with lanthanide contraction, but is different from simple pressure effect. In particular, the $d$-orbitals respond differently in contraction than in pressure.

\section*{Methods}
\subsection*{Calculation details}
All the reported results were obtained with density functional theory (DFT) calculations as implemented in Vienna Abinitio Simulation Package (VASP)\cite{method:vasp,method:pawvasp}. To ensure convergence, we employed plane-wave basis up to 480 eV, and $12\times12\times12$ $\Gamma$-centered K-mesh so that the total energy converges to 1 meV per cell. The Perdew, Burke and Ernzerhoff parameterization (PBE) of generalized gradient approximation (GGA) to the exhange correlation functional was employed\cite{method:pbe}, with which the lattice constants were optimized until the internal stress less than 0.1 kbar. All the electronic structure calculations were obtained with optimized lattice constants unless specified otherwise. Since PBE is known to overestimate the band inversions in solids due to the fact that it overestimates band widths, we have also performed calculations with both modified Becke-Johnson (mBJ) potentials as well as more expensive hybrid functional HSE06 as comparison. Although the rare-earth elements in our calculations have an open $f$-shell, previous studies have shown that in antimonides and bismuthides the $f$-electrons are almost fully localized\cite{PhysRevLett.79.2546,PhysRevB.86.115116,PhysRevB.74.085108,CePn.Pressure.LDAU}, therefore the $f$-electrons are regarded as core state in all calculations. According to photoemission results, most lanthanide elements in LnPn compounds are 3+, therefore we have assumed Ln$^{3+}$ throughout the calculation.

\subsection*{Topological invariant and surface state}
Since LnPn compounds are centrosymmetric, the $\mathcal{Z}_2$ indices can be calculated by evaluating the band parities at 8 time-reversal invariant momenta (TRIM)\cite{TI:z2inv}. The band structures obtained with PBE, mBJ and HSE06 methods were fitted to a tight-binding (TB) Hamiltonian using the maximally localized wannier function (MLWF) method\cite{method:mlwf} with lanthanide-t2g orbitals and pnictogen-$p$ orbitals. The resulting Hamiltonian were then used to calculate the Fermi surfaces as well as surface states using surface Green's function\cite{method:surfgf}.

\section*{Data availability}
The data that support the findings of this study are available from the corresponding author upon reasonable request.

\section*{Acknowledgements}
The authors would like to thank Yi Zhou, Xi Dai, Fuchun Zhang, Jianhui Dai, and Zhi Li for the inspiring discussions. The calculations were partly performed at the Tianhe-2 National Supercomputing Center in China, the HPC center at Hangzhou Normal University, and the HPC at the Center of Correlated Materials in Zhejiang University. This work has been supported by the 973 project (No. 2014CB648400) and the NSFC (No. 11274006, No. 11874137). 

\section*{Author contributions statement}
C.C. conceived and supervised the project; X.D., F.W., J.C, and P.Z. performed calculations; X.D. and C.C. drafted the paper. X.D., F.W., Y.L., H.Y. and C.C. participated in the data analysis and discussion. Xu Duan and Fan Wu contributed equally in this work.

\section*{Additional information}
Correspondence and requests for materials should be addressed to C.C.~(email: ccao@hznu.edu.cn).

\textbf{Competing interests} The authors declare no competing financial or non-financial interests.


\bibliographystyle{naturemag}


\newpage

\newpage

\begin{table}
 \begin{tabular}{c|c|c|c|c|c}
\hline
   1  &  2  & PBE & mBJ & HSE06 & expt \\
  \hline
 \multirow{4}{*}{CeSb} & $\alpha$   & 2.7 (0.23) & 2.4 (0.17) & 2.6 (0.14) & 1.9 (0.23)$^a$ \\
                                        & $\beta_1$ & 7.4 (0.24)  & 5.7 (0.20)  & 6.2 (0.17)  & 3.0 (0.50)$^a$ \\
                                        & $\beta_4$ & 17.3 (0.58) & 14.9 (0.53) & 14.2 (0.46) & 14.0 (4.3)$^d$ \\
                                        & $\gamma$ & 13.9 (0.64) & 10.8 (0.52) & 10.8 (0.44) & 12.2 (0.94)$^a$ \\
  \hline
 \multirow{4}{*}{PrSb} & $\alpha$   & 3.0 (0.22) & 2.6 (0.17) & 2.6 (0.14) & 2.2 (0.21)$^e$ \\
                                        & $\beta_1$ & 8.1 (0.23) & 6.2 (0.21) & 6.2 (0.18) & 4.4$^e$ \\
                                        & $\beta_4$ & 18.4 (0.61) & 15.6 (0.53) & 14.8 (0.43) &  \\
                                        & $\gamma$ & 14.6 (0.63) & 11.4 (0.52) & 11.0 (0.43) & 11.8$^e$ \\
   \hline
 \multirow{4}{*}{SmSb} & $\alpha$   & 3.8 (0.21) & 3.1 (0.17) & 3.1 (0.14) & 3.3 (0.26)$^c$ \\
                                        & $\beta_1$ & 9.1 (0.27) & 6.9 (0.21) & 6.9 (0.18) & 6.3 (0.28)$^c$ \\
                                        & $\beta_4$ & 19.8 (0.57) & 16.5 (0.51) & 15.8 (0.42) & 15.2 (1.4)$^c$ \\
                                        & $\gamma$ & 15.8 (0.57) & 12.0 (0.50) & 11.7 (0.42) & 10.9 (1.3)$^c$ \\
  \hline
 \multirow{4}{*}{CeBi} & $\alpha$   & 3.2 (0.34) & 2.5 (0.24) & 2.4 (0.21) & 2.3 (0.34)$^b$ \\
                                        & $\beta_1$ & 6.7 (0.16) & 6.2 (0.15) & 6.6 (0.14) & 5.6$^b$ \\
                                        & $\beta_4$ & 18.3 (0.59) & 18.5 (0.58) & 18.8 (0.50) & 18.7$^b$  \\
                                        & $\gamma$ & 6.8 (0.77) & 13.1 (0.76) & 13.7 (0.68) & 15.5$^b$  \\
  \hline
 \multirow{4}{*}{PrBi} & $\alpha$   & 7.4 (1.2) & 2.6 (0.28) & 2.7 (0.22) & 2.6 (0.44)$^f$ \\
                                        & $\beta_1$ & 7.2 (0.16) & 6.6 (0.15) & 6.9 (0.14) & 5.5$^f$ \\
                                        & $\beta_4$ & 19.4 (0.60) & 19.2 (0.58) & 19.6 (0.51) & 19.5$^f$ \\
                                        & $\gamma$ & 15.1 (2.0) & 13.6 (0.74) & 14.2 (0.63) & 16.7$^f$ \\
   \hline
 \multirow{4}{*}{SmBi} & $\alpha$   & 3.7 (0.39) & 3.6 (0.23) & 3.7 (0.19) & 3.3 (0.44)$^f$ \\
                                        & $\beta_1$ & 8.3 (0.18) & 7.3 (0.16) & 7.6 (0.14) & 6.7$^f$  \\
                                        & $\beta_4$ & 22.7 (0.65) & 20.9 (0.57) & 20.9 (0.49) &  \\
                                        & $\gamma$ & 17.5 (0.86) & 14.7 (0.62) & 14.9 (0.52) &14.6$^f$ \\
  \hline
\end{tabular}
 \caption{Calculated de Haas van Alphen (dHvA)/Shubnikov de Haas (SdH) frequencies (in kT) ($\mathbf{B}\parallel$ [001]) compared with experimental observations. The numbers in the brackets are the effective mass $\vert m*\vert$ in $m_e$. The $\alpha$ and $\gamma$ bands are the electron pockets near $X$, whereas the $\beta_1$ and $\beta_4$ bands are hole-type pockets near $\Gamma$. The experimental values were extracted from $^a$ Ref. \cite{dHvA.CeSb}, $^b$ Ref. \cite{MORITA1997192}, $^c$ Ref. \cite{OZEKI1991499}, $^d$ Ref. \cite{CeSb.Heavy.Hole.dHvA}, $^e$ Ref. \cite{PhysRevB.96.125122}, and $^f$ unpublished SdH results. \label{tab:dhva}}
\end{table}

\end{document}